%Paper: hep-lat/9208003
%From: Pablo Tamayo <tamayo@Think.COM>
%Date: Mon, 3 Aug 92 19:26:26 EDT
%Date (revised): Wed, 5 Aug 92 12:19:01 EDT

% This is a TeX file
% The 10 PostScript figures are available by anonymous FTP from think.com:
%
%  ftp think.com    cd users/tamayo    get cluster_figs.ps  (1.38 Mbytes)
%
%
% VANILLA.STY
% COPYRIGHT (C) 1985, 1986 BY MICHAEL SPIVAK
% version date 1/1/86
\catcode`\@=11
\font\tensmc=cmcsc10      %change to CM fonts 3-31-87
%\font\tensmc=amcsc10
\def\smc{\tensmc}

\def\hcorrection#1{\advance\hoffset by #1 }
\def\vcorrection#1{\advance\voffset by #1 }
\def\wlog#1{}
\newif\iftitle@
\outer\def\title{\title@true\vglue 24\p@ plus 12\p@ minus 12\p@
   \bgroup\let\\=\cr\tabskip\centering
   \halign to \hsize\bgroup\tenbf\hfill\ignorespaces##\unskip\hfill\cr}
\def\endtitle{\cr\egroup\egroup\vglue 18\p@ plus 12\p@ minus 6\p@}
\outer\def\author{\iftitle@\vglue -18\p@ plus -12\p@ minus -6\p@\fi\vglue
    12\p@ plus 6\p@ minus 3\p@\bgroup\let\\=\cr\tabskip\centering
    \halign to \hsize\bgroup\smc\hfill\ignorespaces##\unskip\hfill\cr}
\def\endauthor{\cr\egroup\egroup\vglue 18\p@ plus 12\p@ minus 6\p@}
\outer\def\heading{\bigbreak\bgroup\let\\=\cr\tabskip\centering
    \halign to \hsize\bgroup\smc\hfill\ignorespaces##\unskip\hfill\cr}
\def\endheading{\cr\egroup\egroup\nobreak\medskip}

\outer\def\endproclaim{\par\ifdim\lastskip<\medskipamount\removelastskip
  \penalty 55 \fi\medskip\rm}
\outer\def\demo#1{\par\ifdim\lastskip<\smallskipamount\removelastskip
    \smallskip\fi\noindent{\smc\ignorespaces#1\unskip:\enspace}\rm
      \ignorespaces}

\newcount\footmarkcount@
\footmarkcount@=1
\def\makefootnote@#1#2{\insert\footins{\interlinepenalty=100
  \splittopskip=\ht\strutbox \splitmaxdepth=\dp\strutbox
  \floatingpenalty=\@MM
  \leftskip=\z@\rightskip=\z@\spaceskip=\z@\xspaceskip=\z@
  \noindent{#1}\footstrut\rm\ignorespaces #2\strut}}
\def\footnote{\let\@sf=\empty\ifhmode\edef\@sf{\spacefactor
   =\the\spacefactor}\/\fi\futurelet\next\footnote@}
\def\footnote@{\ifx"\next\let\next\footnote@@\else
    \let\next\footnote@@@\fi\next}
\def\footnote@@"#1"#2{#1\@sf\relax\makefootnote@{#1}{#2}}
\def\footnote@@@#1{${\number\footmarkcount@}$\makefootnote@
   {${\number\footmarkcount@}$}{#1}\global\advance\footmarkcount@ by 1 }

\hyphenation{man-u-script man-u-scripts ap-pen-dix ap-pen-di-ces}
\hyphenation{data-base data-bases}
\ifx\amstexloaded@\relax\catcode`\@=13
  \endinput\else\let\amstexloaded@=\relax\fi
\newlinechar=`\
\def\eat@#1{}
\def\Space@.{\futurelet\Space@\relax}
\Space@. %
\newhelp\athelp@
{Only certain combinations beginning with @ make sense to me.
Perhaps you wanted \string\@\space for a printed @?
I've ignored the character or group after @.}
\def\futureletnextat@{\futurelet\next\at@}
{\catcode`\@=\active
\lccode`\Z=`\@ \lowercase
{\gdef@{\expandafter\csname futureletnextatZ\endcsname}
\expandafter\gdef\csname atZ\endcsname
   {\ifcat\noexpand\next a\def\next{\csname atZZ\endcsname}\else
   \ifcat\noexpand\next0\def\next{\csname atZZ\endcsname}\else
    \def\next{\csname atZZZ\endcsname}\fi\fi\next}
\expandafter\gdef\csname atZZ\endcsname#1{\expandafter
   \ifx\csname #1Zat\endcsname\relax\def\next
     {\errhelp\expandafter=\csname athelpZ\endcsname
      \errmessage{Invalid use of \string@}}\else
       \def\next{\csname #1Zat\endcsname}\fi\next}
\expandafter\gdef\csname atZZZ\endcsname#1{\errhelp
    \expandafter=\csname athelpZ\endcsname
      \errmessage{Invalid use of \string@}}}}
\def\atdef@#1{\expandafter\def\csname #1@at\endcsname}
\newhelp\defahelp@{If you typed \string\define\space cs instead of
\string\define\string\cs\space
I've substituted an inaccessible control sequence so that your
definition will be completed without mixing me up too badly.
If you typed \string\define{\string\cs} the inaccessible control sequence
was defined to be \string\cs, and the rest of your
definition appears as input.}
\newhelp\defbhelp@{I've ignored your definition, because it might
conflict with other uses that are important to me.}
\def\define{\futurelet\next\define@}
\def\define@{\ifcat\noexpand\next\relax
  \def\next{\define@@}%
  \else\errhelp=\defahelp@
  \errmessage{\string\define\space must be followed by a control
     sequence}\def\next{\def\garbage@}\fi\next}
\def\undefined@{}
\def\preloaded@{}
\def\define@@#1{\ifx#1\relax\errhelp=\defbhelp@
   \errmessage{\string#1\space is already defined}\def\next{\def\garbage@}%
   \else\expandafter\ifx\csname\expandafter\eat@\string
         #1@\endcsname\undefined@\errhelp=\defbhelp@
   \errmessage{\string#1\space can't be defined}\def\next{\def\garbage@}%
   \else\expandafter\ifx\csname\expandafter\eat@\string#1\endcsname\relax
     \def\next{\def#1}\else\errhelp=\defbhelp@
     \errmessage{\string#1\space is already defined}\def\next{\def\garbage@}%
      \fi\fi\fi\next}
\def\famzero{\fam\z@}

\def\lim{\mathop{\famzero lim}}

\def\log{\mathop{\famzero log}\nolimits}

\def\textfont@#1#2{\def#1{\relax\ifmmode
    \errmessage{Use \string#1\space only in text}\else#2\fi}}
\textfont@\rm\tenrm
\textfont@\it\tenit
\textfont@\sl\tensl
\textfont@\bf\tenbf
\textfont@\smc\tensmc
\let\ic@=\/
\def\/{\unskip\ic@}
\def\textfonti{\the\textfont1 }
\def\t#1#2{{\edef\next{\the\font}\textfonti\accent"7F \next#1#2}}
\let\B=\=
\let\D=\.
\def~{\unskip\nobreak\ \ignorespaces}
{\catcode`\@=\active
\gdef\@{\char'100 }}
\atdef@-{\leavevmode\futurelet\next\athyph@}
\def\athyph@{\ifx\next-\let\next=\athyph@@
  \else\let\next=\athyph@@@\fi\next}
\def\athyph@@@{\hbox{-}}
\def\athyph@@#1{\futurelet\next\athyph@@@@}
\def\athyph@@@@{\if\next-\def\next##1{\hbox{---}}\else
    \def\next{\hbox{--}}\fi\next}
\def\.{.\spacefactor=\@m}
\atdef@.{\null.}
\atdef@,{\null,}
\atdef@;{\null;}
\atdef@:{\null:}
\atdef@?{\null?}
\atdef@!{\null!}
\def\srdr@{\thinspace}
\def\drsr@{\kern.02778em}
\def\sldl@{\kern.02778em}
\def\dlsl@{\thinspace}
\atdef@"{\unskip\futurelet\next\atqq@}
\def\atqq@{\ifx\next\Space@\def\next. {\atqq@@}\else
         \def\next.{\atqq@@}\fi\next.}
\def\atqq@@{\futurelet\next\atqq@@@}
\def\atqq@@@{\ifx\next`\def\next`{\atqql@}\else\def\next'{\atqqr@}\fi\next}
\def\atqql@{\futurelet\next\atqql@@}
\def\atqql@@{\ifx\next`\def\next`{\sldl@``}\else\def\next{\dlsl@`}\fi\next}
\def\atqqr@{\futurelet\next\atqqr@@}
\def\atqqr@@{\ifx\next'\def\next'{\srdr@''}\else\def\next{\drsr@'}\fi\next}

\def\textfontii{\the\textfont2 }
\def\{{\relax\ifmmode\lbrace\else
    {\textfontii f}\spacefactor=\@m\fi}
\def\}{\relax\ifmmode\rbrace\else
    \let\@sf=\empty\ifhmode\edef\@sf{\spacefactor=\the\spacefactor}\fi
      {\textfontii g}\@sf\relax\fi}
\def\nonhmodeerr@#1{\errmessage
     {\string#1\space allowed only within text}}
\def\linebreak{\relax\ifhmode\unskip\break\else
    \nonhmodeerr@\linebreak\fi}
\def\allowlinebreak{\relax
   \ifhmode\allowbreak\else\nonhmodeerr@\allowlinebreak\fi}
\newskip\saveskip@
\def\nolinebreak{\relax\ifhmode\saveskip@=\lastskip\unskip
  \nobreak\ifdim\saveskip@>\z@\hskip\saveskip@\fi
   \else\nonhmodeerr@\nolinebreak\fi}
\def\newline{\relax\ifhmode\null\hfil\break
    \else\nonhmodeerr@\newline\fi}
\def\nonmathaerr@#1{\errmessage
     {\string#1\space is not allowed in display math mode}}
\def\nonmathberr@#1{\errmessage{\string#1\space is allowed only in math mode}}
\def\mathbreak{\relax\ifmmode\ifinner\break\else
   \nonmathaerr@\mathbreak\fi\else\nonmathberr@\mathbreak\fi}
\def\nomathbreak{\relax\ifmmode\ifinner\nobreak\else
    \nonmathaerr@\nomathbreak\fi\else\nonmathberr@\nomathbreak\fi}
\def\allowmathbreak{\relax\ifmmode\ifinner\allowbreak\else
     \nonmathaerr@\allowmathbreak\fi\else\nonmathberr@\allowmathbreak\fi}
\def\pagebreak{\relax\ifmmode
   \ifinner\errmessage{\string\pagebreak\space
     not allowed in non-display math mode}\else\postdisplaypenalty-\@M\fi
   \else\ifvmode\penalty-\@M\else\edef\spacefactor@
       {\spacefactor=\the\spacefactor}\vadjust{\penalty-\@M}\spacefactor@
        \relax\fi\fi}
\def\nopagebreak{\relax\ifmmode
     \ifinner\errmessage{\string\nopagebreak\space
    not allowed in non-display math mode}\else\postdisplaypenalty\@M\fi
    \else\ifvmode\nobreak\else\edef\spacefactor@
        {\spacefactor=\the\spacefactor}\vadjust{\penalty\@M}\spacefactor@
         \relax\fi\fi}
\def\newpage{\relax\ifvmode\vfill\penalty-\@M\else\nonvmodeerr@\newpage\fi}
\def\nonvmodeerr@#1{\errmessage
    {\string#1\space is allowed only between paragraphs}}
\def\smallpagebreak{\relax\ifvmode\smallbreak
      \else\nonvmodeerr@\smallpagebreak\fi}
\def\medpagebreak{\relax\ifvmode\medbreak
       \else\nonvmodeerr@\medpagebreak\fi}
\def\bigpagebreak{\relax\ifvmode\bigbreak
      \else\nonvmodeerr@\bigpagebreak\fi}
\newdimen\captionwidth@
\captionwidth@=\hsize
\advance\captionwidth@ by -1.5in
\def\caption#1{}
\def\topspace#1{\gdef\thespace@{#1}\ifvmode\def\next
    {\futurelet\next\topspace@}\else\def\next{\nonvmodeerr@\topspace}\fi\next}
\def\topspace@{\ifx\next\Space@\def\next. {\futurelet\next\topspace@@}\else
     \def\next.{\futurelet\next\topspace@@}\fi\next.}
\def\topspace@@{\ifx\next\caption\let\next\topspace@@@\else
    \let\next\topspace@@@@\fi\next}
 \def\topspace@@@@{\topinsert\vbox to
       \thespace@{}\endinsert}
\def\topspace@@@\caption#1{\topinsert\vbox to
    \thespace@{}\nobreak
      \smallskip
    \setbox\z@=\hbox{\noindent\ignorespaces#1\unskip}%
   \ifdim\wd\z@>\captionwidth@
   \centerline{\vbox{\hsize=\captionwidth@\noindent\ignorespaces#1\unskip}}%
   \else\centerline{\box\z@}\fi\endinsert}
\def\midspace#1{\gdef\thespace@{#1}\ifvmode\def\next
    {\futurelet\next\midspace@}\else\def\next{\nonvmodeerr@\midspace}\fi\next}
\def\midspace@{\ifx\next\Space@\def\next. {\futurelet\next\midspace@@}\else
     \def\next.{\futurelet\next\midspace@@}\fi\next.}
\def\midspace@@{\ifx\next\caption\let\next\midspace@@@\else
    \let\next\midspace@@@@\fi\next}
 \def\midspace@@@@{\midinsert\vbox to
       \thespace@{}\endinsert}
\def\midspace@@@\caption#1{\midinsert\vbox to
    \thespace@{}\nobreak
      \smallskip
      \setbox\z@=\hbox{\noindent\ignorespaces#1\unskip}%
      \ifdim\wd\z@>\captionwidth@
    \centerline{\vbox{\hsize=\captionwidth@\noindent\ignorespaces#1\unskip}}%
    \else\centerline{\box\z@}\fi\endinsert}
\mathchardef\prime@="0230
\def\prime{{{}\prime@{}}}
\def\prim@s{\prime@\futurelet\next\pr@m@s}

\def\,{\relax\ifmmode\mskip\thinmuskip\else\thinspace\fi}
\def\!{\relax\ifmmode\mskip-\thinmuskip\else\negthinspace\fi}
\def\frac#1#2{{#1\over#2}}

\def\:{\nobreak\hskip.1111em{:}\hskip.3333em plus .0555em\relax}
\def\intic@{\mathchoice{\hskip5\p@}{\hskip4\p@}{\hskip4\p@}{\hskip4\p@}}
\def\negintic@
 {\mathchoice{\hskip-5\p@}{\hskip-4\p@}{\hskip-4\p@}{\hskip-4\p@}}
\def\intkern@{\mathchoice{\!\!\!}{\!\!}{\!\!}{\!\!}}
\def\intdots@{\mathchoice{\cdots}{{\cdotp}\mkern1.5mu
    {\cdotp}\mkern1.5mu{\cdotp}}{{\cdotp}\mkern1mu{\cdotp}\mkern1mu
      {\cdotp}}{{\cdotp}\mkern1mu{\cdotp}\mkern1mu{\cdotp}}}
\newcount\intno@
\def\iint{\intno@=\tw@\futurelet\next\ints@}
\def\iiint{\intno@=\thr@@\futurelet\next\ints@}
\def\iiiint{\intno@=4 \futurelet\next\ints@}
\def\idotsint{\intno@=\z@\futurelet\next\ints@}
\def\ints@{\findlimits@\ints@@}
\newif\iflimtoken@
\newif\iflimits@
\def\findlimits@{\limtoken@false\limits@false\ifx\next\limits
 \limtoken@true\limits@true\else\ifx\next\nolimits\limtoken@true\limits@false
    \fi\fi}
\def\multintlimits@{\intop\ifnum\intno@=\z@\intdots@
  \else\intkern@\fi
    \ifnum\intno@>\tw@\intop\intkern@\fi
     \ifnum\intno@>\thr@@\intop\intkern@\fi\intop}
\def\multint@{\int\ifnum\intno@=\z@\intdots@\else\intkern@\fi
   \ifnum\intno@>\tw@\int\intkern@\fi
    \ifnum\intno@>\thr@@\int\intkern@\fi\int}
\def\ints@@{\iflimtoken@\def\ints@@@{\iflimits@
   \negintic@\mathop{\intic@\multintlimits@}\limits\else
    \multint@\nolimits\fi\eat@}\else
     \def\ints@@@{\multint@\nolimits}\fi\ints@@@}
\def\Sb{_\bgroup\vspace@
        \baselineskip=\fontdimen10 \scriptfont\tw@
        \advance\baselineskip by \fontdimen12 \scriptfont\tw@
        \lineskip=\thr@@\fontdimen8 \scriptfont\thr@@
        \lineskiplimit=\thr@@\fontdimen8 \scriptfont\thr@@
        \Let@\vbox\bgroup\halign\bgroup \hfil$\scriptstyle
            {##}$\hfil\cr}
\def\endSb{\crcr\egroup\egroup\egroup}
\def\Sp{\bgroup\vspace@
        \baselineskip=\fontdimen10 \scriptfont\tw@
        \advance\baselineskip by \fontdimen12 \scriptfont\tw@
        \lineskip=\thr@@\fontdimen8 \scriptfont\thr@@
        \lineskiplimit=\thr@@\fontdimen8 \scriptfont\thr@@
        \Let@\vbox\bgroup\halign\bgroup \hfil$\scriptstyle
            {##}$\hfil\cr}
\def\endSp{\crcr\egroup\egroup\egroup}
\def\Let@{\relax\iffalse{\fi\let\\=\cr\iffalse}\fi}
\def\vspace@{\def\vspace##1{\noalign{\vskip##1 }}}
\def\aligned{\,\vcenter\bgroup\vspace@\Let@\openup\jot\m@th\ialign
  \bgroup \strut\hfil$\displaystyle{##}$&$\displaystyle{{}##}$\hfil\crcr}
\def\endaligned{\crcr\egroup\egroup}
\def\matrix{\,\vcenter\bgroup\Let@\vspace@
    \normalbaselines
  \m@th\ialign\bgroup\hfil$##$\hfil&&\quad\hfil$##$\hfil\crcr
    \mathstrut\crcr\noalign{\kern-\baselineskip}}
\def\endmatrix{\crcr\mathstrut\crcr\noalign{\kern-\baselineskip}\egroup
                \egroup\,}
\newtoks\hashtoks@
\hashtoks@={#}
\def\format{\crcr\egroup\iffalse{\fi\ifnum`}=0 \fi\format@}
\def\format@#1\\{\def\preamble@{#1}%
  \def\c{\hfil$\the\hashtoks@$\hfil}%
  \def\r{\hfil$\the\hashtoks@$}%
  \def\l{$\the\hashtoks@$\hfil}%
  \setbox\z@=\hbox{\xdef\Preamble@{\preamble@}}\ifnum`{=0 \fi\iffalse}\fi
   \ialign\bgroup\span\Preamble@\crcr}

\def\cases{\left\{\,\vcenter\bgroup\vspace@
     \normalbaselines\openup\jot\m@th
       \Let@\ialign\bgroup$##$\hfil&\quad$##$\hfil\crcr
      \mathstrut\crcr\noalign{\kern-\baselineskip}}

\newif\iftagsleft@
\tagsleft@true
\def\TagsOnRight{\global\tagsleft@false}
\def\tag#1$${\iftagsleft@\leqno\else\eqno\fi
 \hbox{\def\pagebreak{\global\postdisplaypenalty-\@M}%
 \def\nopagebreak{\global\postdisplaypenalty\@M}\rm(#1\unskip)}%
  $$\postdisplaypenalty\z@\ignorespaces}
\interdisplaylinepenalty=\@M
\def\allowdisplaybreak@{\def\allowdisplaybreak{\noalign{\allowbreak}}}
\def\displaybreak@{\def\displaybreak{\noalign{\break}}}
\def\align#1\endalign{\def\tag{&}\vspace@\allowdisplaybreak@\displaybreak@
  \iftagsleft@\lalign@#1\endalign\else
   \ralign@#1\endalign\fi}
\def\ralign@#1\endalign{\displ@y\Let@\tabskip\centering\halign to\displaywidth
     {\hfil$\displaystyle{##}$\tabskip=\z@&$\displaystyle{{}##}$\hfil
       \tabskip=\centering&\llap{\hbox{(\rm##\unskip)}}\tabskip\z@\crcr
             #1\crcr}}
\def\lalign@
 #1\endalign{\displ@y\Let@\tabskip\centering\halign to \displaywidth
   {\hfil$\displaystyle{##}$\tabskip=\z@&$\displaystyle{{}##}$\hfil
   \tabskip=\centering&\kern-\displaywidth
        \rlap{\hbox{(\rm##\unskip)}}\tabskip=\displaywidth\crcr
               #1\crcr}}
\def\overrightarrow{\mathpalette\overrightarrow@}
\def\overrightarrow@#1#2{\vbox{\ialign{$##$\cr
    #1{-}\mkern-6mu\cleaders\hbox{$#1\mkern-2mu{-}\mkern-2mu$}\hfill
     \mkern-6mu{\to}\cr
     \noalign{\kern -1\p@\nointerlineskip}
     \hfil#1#2\hfil\cr}}}
\def\overleftarrow{\mathpalette\overleftarrow@}
\def\overleftarrow@#1#2{\vbox{\ialign{$##$\cr
     #1{\leftarrow}\mkern-6mu\cleaders\hbox{$#1\mkern-2mu{-}\mkern-2mu$}\hfill
      \mkern-6mu{-}\cr
     \noalign{\kern -1\p@\nointerlineskip}
     \hfil#1#2\hfil\cr}}}
\def\overleftrightarrow{\mathpalette\overleftrightarrow@}
\def\overleftrightarrow@#1#2{\vbox{\ialign{$##$\cr
     #1{\leftarrow}\mkern-6mu\cleaders\hbox{$#1\mkern-2mu{-}\mkern-2mu$}\hfill
       \mkern-6mu{\to}\cr
    \noalign{\kern -1\p@\nointerlineskip}
      \hfil#1#2\hfil\cr}}}
\def\underrightarrow{\mathpalette\underrightarrow@}
\def\underrightarrow@#1#2{\vtop{\ialign{$##$\cr
    \hfil#1#2\hfil\cr
     \noalign{\kern -1\p@\nointerlineskip}
    #1{-}\mkern-6mu\cleaders\hbox{$#1\mkern-2mu{-}\mkern-2mu$}\hfill
     \mkern-6mu{\to}\cr}}}
\def\underleftarrow{\mathpalette\underleftarrow@}
\def\underleftarrow@#1#2{\vtop{\ialign{$##$\cr
     \hfil#1#2\hfil\cr
     \noalign{\kern -1\p@\nointerlineskip}
     #1{\leftarrow}\mkern-6mu\cleaders\hbox{$#1\mkern-2mu{-}\mkern-2mu$}\hfill
      \mkern-6mu{-}\cr}}}
\def\underleftrightarrow{\mathpalette\underleftrightarrow@}
\def\underleftrightarrow@#1#2{\vtop{\ialign{$##$\cr
      \hfil#1#2\hfil\cr
    \noalign{\kern -1\p@\nointerlineskip}
     #1{\leftarrow}\mkern-6mu\cleaders\hbox{$#1\mkern-2mu{-}\mkern-2mu$}\hfill
       \mkern-6mu{\to}\cr}}}
\def\sqrt#1{\radical"270370 {#1}}
\def\dots{\relax\ifmmode\let\next=\ldots\else\let\next=\tdots@\fi\next}
\def\tdots@{\unskip\ \tdots@@}
\def\tdots@@{\futurelet\next\tdots@@@}
\def\tdots@@@{$\mathinner{\ldotp\ldotp\ldotp}\,
   \ifx\next,$\else
   \ifx\next.\,$\else
   \ifx\next;\,$\else
   \ifx\next:\,$\else
   \ifx\next?\,$\else
   \ifx\next!\,$\else
   $ \fi\fi\fi\fi\fi\fi}
\def\text{\relax\ifmmode\let\next=\text@\else\let\next=\text@@\fi\next}
\def\text@@#1{\hbox{#1}}
\def\text@#1{\mathchoice
 {\hbox{\everymath{\displaystyle}\def\textfonti{\the\textfont1 }%
    \def\textfontii{\the\textfont2 }\textdef@@ T#1}}
 {\hbox{\everymath{\textstyle}\def\textfonti{\the\textfont1 }%
    \def\textfontii{\the\textfont2 }\textdef@@ T#1}}
 {\hbox{\everymath{\scriptstyle}\def\textfonti{\the\scriptfont1 }%
   \def\textfontii{\the\scriptfont2 }\textdef@@ S\rm#1}}
 {\hbox{\everymath{\scriptscriptstyle}\def\textfonti{\the\scriptscriptfont1 }%
   \def\textfontii{\the\scriptscriptfont2 }\textdef@@ s\rm#1}}}
\def\textdef@@#1{\textdef@#1\rm \textdef@#1\bf
   \textdef@#1\sl \textdef@#1\it}

\def\textdef@#1#2{\def\next{\csname\expandafter\eat@\string#2fam\endcsname}%
\if S#1\edef#2{\the\scriptfont\next\relax}%
 \else\if s#1\edef#2{\the\scriptscriptfont\next\relax}%
 \else\edef#2{\the\textfont\next\relax}\fi\fi}
\scriptfont\itfam=\tenit \scriptscriptfont\itfam=\tenit
\scriptfont\slfam=\tensl \scriptscriptfont\slfam=\tensl
\mathcode`\0="0030
\mathcode`\1="0031
\mathcode`\2="0032
\mathcode`\3="0033
\mathcode`\4="0034
\mathcode`\5="0035
\mathcode`\6="0036
\mathcode`\7="0037
\mathcode`\8="0038
\mathcode`\9="0039
\def\Cal{\relax\ifmmode\let\next=\Cal@\else
     \def\next{\errmessage{Use \string\Cal\space only in math mode}}\fi\next}
\def\Cal@#1{{\fam2 #1}}
\def\bold{\relax\ifmmode\let\next=\bold@\else
   \def\next{\errmessage{Use \string\bold\space only in math
      mode}}\fi\next}\def\bold@#1{{\fam\bffam #1}}
\mathchardef\Gamma="0000
\mathchardef\Delta="0001
\mathchardef\Theta="0002
\mathchardef\Lambda="0003
\mathchardef\Xi="0004
\mathchardef\Pi="0005
\mathchardef\Sigma="0006
\mathchardef\Upsilon="0007
\mathchardef\Phi="0008
\mathchardef\Psi="0009
\mathchardef\Omega="000A
\mathchardef\varGamma="0100
\mathchardef\varDelta="0101
\mathchardef\varTheta="0102
\mathchardef\varLambda="0103
\mathchardef\varXi="0104
\mathchardef\varPi="0105
\mathchardef\varSigma="0106
\mathchardef\varUpsilon="0107
\mathchardef\varPhi="0108
\mathchardef\varPsi="0109
\mathchardef\varOmega="010A
\font\dummyft@=dummy
\fontdimen1 \dummyft@=\z@
\fontdimen2 \dummyft@=\z@
\fontdimen3 \dummyft@=\z@
\fontdimen4 \dummyft@=\z@
\fontdimen5 \dummyft@=\z@
\fontdimen6 \dummyft@=\z@
\fontdimen7 \dummyft@=\z@
\fontdimen8 \dummyft@=\z@
\fontdimen9 \dummyft@=\z@
\fontdimen10 \dummyft@=\z@
\fontdimen11 \dummyft@=\z@
\fontdimen12 \dummyft@=\z@
\fontdimen13 \dummyft@=\z@
\fontdimen14 \dummyft@=\z@
\fontdimen15 \dummyft@=\z@
\fontdimen16 \dummyft@=\z@
\fontdimen17 \dummyft@=\z@
\fontdimen18 \dummyft@=\z@
\fontdimen19 \dummyft@=\z@
\fontdimen20 \dummyft@=\z@
\fontdimen21 \dummyft@=\z@
\fontdimen22 \dummyft@=\z@
\def\fontlist@{\\{\tenrm}\\{\sevenrm}\\{\fiverm}\\{\teni}\\{\seveni}%
 \\{\fivei}\\{\tensy}\\{\sevensy}\\{\fivesy}\\{\tenex}\\{\tenbf}\\{\sevenbf}%
 \\{\fivebf}\\{\tensl}\\{\tenit}\\{\tensmc}}
\def\dodummy@{{\def\\##1{\global\let##1=\dummyft@}\fontlist@}}
\newif\ifsyntax@
\newcount\countxviii@
\def\newtoks@{\alloc@5\toks\toksdef\@cclvi}
\def\nopages@{\output={\setbox\z@=\box\@cclv \deadcycles=\z@}\newtoks@\output}
\def\syntax{\syntax@true\dodummy@\countxviii@=\count18
\loop \ifnum\countxviii@ > \z@ \textfont\countxviii@=\dummyft@
   \scriptfont\countxviii@=\dummyft@ \scriptscriptfont\countxviii@=\dummyft@
     \advance\countxviii@ by-\@ne\repeat
\dummyft@\tracinglostchars=\z@
  \nopages@\frenchspacing\hbadness=\@M}
\def\magstep#1{\ifcase#1 1000\or
 1200\or 1440\or 1728\or 2074\or 2488\or
 \errmessage{\string\magstep\space only works up to 5}\fi\relax}
{\lccode`\2=`\p \lccode`\3=`\t
 \lowercase{\gdef\tru@#123{#1truept}}}

\def\scaletype#1{\mag=#1\relax
 \hsize=\expandafter\tru@\the\hsize
 \vsize=\expandafter\tru@\the\vsize
 \dimen\footins=\expandafter\tru@\the\dimen\footins}

\def\scalefont#1#2\andcallit#3{\edef\font@{\the\font}#1\font#3=
  \fontname\font\space scaled #2\relax\font@}
\def\Mag@#1#2{\ifdim#1<1pt\multiply#1 #2\relax\divide#1 1000 \else
  \ifdim#1<10pt\divide#1 10 \multiply#1 #2\relax\divide#1 100\else
  \divide#1 100 \multiply#1 #2\relax\divide#1 10 \fi\fi}
\def\scalelinespacing#1{\Mag@\baselineskip{#1}\Mag@\lineskip{#1}%
  \Mag@\lineskiplimit{#1}}
\def\wlog#1{\immediate\write-1{#1}}
\catcode`\@=\active

{}.

\font\tenbf=cmbx10
\font\tenrm=cmr10
\font\tenit=cmti10
\font\ninebf=cmbx9
\font\ninerm=cmr9
\font\nineit=cmti9

\font\eightrm=cmr8
\font\eightit=cmti8
\font\sevenrm=cmr7
\TagsOnRight
\hsize=5.0truein
\vsize=7.7truein
\parindent=15pt
\nopagenumbers
\baselineskip=10pt
\centerline{\eightrm   }
\centerline{\eightrm   }

\vglue 5pc
\baselineskip=13pt
\centerline{\tenbf A PARALLEL CLUSTER LABELING METHOD}
\centerline{\tenbf FOR MONTE CARLO DYNAMICS}
\vglue 1cm
\centerline{\tenrm Mike Flanigan and Pablo Tamayo}
\baselineskip=12pt
\centerline{\eightit Thinking Machines Corp.}
\baselineskip=10pt
\centerline{\eightit 245 First. St. Cambridge, MA 02142 U.S.A}
\vglue 1cm
\centerline{\eightrm Received (........)}
\baselineskip=10pt
\centerline{\eightrm Revised (........)}
\vglue 0.5cm

\centerline{\eightrm ABSTRACT}
{\rightskip=1.5pc \leftskip=1.5pc \eightrm\baselineskip=10pt\noindent We
present an algorithm for cluster dynamics to efficiently simulate large
systems on MIMD parallel computers with large numbers of processing nodes.
The method divides physical space into rectangular cells which are assigned
to processing nodes and combines a serial procedure, by which clusters are
labeled locally inside each cell, with a nearest neighbor relaxation
process in which processing nodes exchange labels until a fixed point is
reached.  By controlling overhead and reducing inter-processor
communication this method attains good performance and speed-up. The
complexity and scaling properties of the algorithm are analyzed.  The
algorithm has been used to simulate large two-dimensional Ising systems (up
to 27808 $\times$ 27808 sites) with Swendsen-Wang dynamics.  Typical
updating times on the order of 82 nanosecs/site and efficiencies larger
than 90\% have been obtained using 256 processing nodes on the CM-5
supercomputer.
\vglue 0.3cm
\noindent
{\eightit Keywords}\/: Ising Model, Cluster Labeling, Percolation, Monte Carlo
Simulations, Accelerated Dynamics.
\vglue 0.5cm}
\baselineskip=13pt

\vglue 16pt
\line{\tenbf 1. Introduction\hfil}
\vglue 5pt

Over the last five years, since the introduction of the Swendsen-Wang (SW)
algorithm,$^1$ cluster dynamics have been extended to a variety of models
in Statistical Mechanics and Field Theory$^2$. These new ``accelerated
dynamics'' update percolation clusters instead of single spins producing
faster decorrelation times and reduced critical slowing down$^3$. In fact,
the dynamic critical exponent $z$ is reduced significantly or even
eliminated in some cases. This acceleration does not come for free; the
price to pay is an increased computational complexity compared with
standard Metropolis, heat bath or other local algorithms. The cluster
labeling procedure requires a fair amount of non-local (unstructured) data
movements that makes cluster algorithms intrinsically hard to vectorize or
paralellize. It is important to develop efficient cluster labeling methods
because they are used every time step inside the core of the simulations.
One has to worry not only about their complexity and scaling properties but
also about absolute execution times. One of the modern challenges of
computational science is to find efficient algorithms to exploit the
unprecedented computational power of today's massively parallel
supercomputers.

Cluster labeling on a lattice is also relevant for the analysis of
structures obtained in many computer simulations of statistical systems:
Ising and percolation clusters,$^4$ nucleation droplets, polymers,
crystals, fractal structures, and particle tracks in experimental high
energy Physics. Cluster labeling is a special case of the more general
problem of finding the connected components of a graph,$^5$ which has
applications in computer vision, image processing and network analysis,
among others.

\headline={\ifodd\pageno\rightheadline \else\leftheadline\fi}
\def\rightheadline{\eightit\hfil A Parallel Cluster Labeling
Method for Monte Carlo Dynamics\quad
\eightrm\folio}
\def\leftheadline{\eightrm\folio\quad\eightit M. Flanigan \&
 P. Tamayo \hfil}
\voffset=2\baselineskip

In this paper we propose a simple method for cluster labeling that can be
used at the core of Monte Carlo simulations on
MIMD\footnote"$^i$"{\eightrm\baselineskip=10pt Multiple Instruction
Multiple Data} parallel computers and in particular the CM-5 (Thinking
Machines Corp.).  The algorithm we will present is a general method for
cluster finding in $n-$dimensional Euclidean lattices, but we will
concentrate on the specific problem of cluster labeling for the Ising Model
with Swendsen-Wang dynamics.  Percolation bonds are defined between aligned
spins with bond probability $p_{bond} = 1 - e^{-2\beta}$, and the clusters
of connected spins, the Coniglio-Klein$^{6,7,8}$ percolation clusters, are
flipped with 50\% probability. At the critical point, where most of the
interesting behavior occurs, the clusters span the system and information
has to be propagated across the entire computational domain.

\midinsert\vskip 9cm
{\rightskip=1.5pc \leftskip=1.5pc \eightrm\baselineskip=10pt\noindent
Fig. 1.\,\,\, Typical critical clusters in the
two-dimensional Ising model. The lines show the partitioning of the system
into 16 cells (processing nodes).}
\endinsert

Several methods to find connected components have been introduced in the
Computer Science literature. There are general methods based on
union-find,$^{9,10}$ transitive closure,$^{11}$ vertex
collapse$^{11,12,13,14}$ and vector models$^{15}$. We realized that most of
these methods are not well suited to practical lattice Monte Carlo
simulations on MIMD machines where one has to worry about absolute
execution times over average configurations instead of worse cases. Most of
these methods are designed for idealized
PRAM\footnote"$^{ii}$"{\eightrm\baselineskip=10pt Parallel Random-Access
Machine.} models and require extensive global communications, shared memory
or data format transformations. Some other methods are more appropriate for
SIMD\footnote"$^{iii}$"{\eightrm\baselineskip=10pt Single Instruction
Multiple Data.} fine-grained machines$^{15,16,17,25,29}$. The choice of
data partitioning is very important$^{18,9}$. A number of MIMD cluster
labeling algorithms, specifically designed for cluster Monte Carlo
simulations have appeared in the literature,$^{19,21}$ but have very
limited speed-up and efficiency which limits their application to the
simulation of large systems using large numbers of processing nodes. The
algorithm with the best scaling properties appears to be the self-labeling
method introduced by Baillie and Coddington$^{21}$.  Recently, Kertesz and
Stauffer$^{20}$ have used the strip geometric parallelization method to
simulate large systems ($6400^2$) on the Intel iPSC/860.  These methods
have greatly improved our knowledge of the behavior of large systems but
still lack the required scaling properties to efficiently utilize hundreds
or thousands of processing nodes. In this paper we propose a method
appropriate for the simulation of large systems on large numbers of
processing nodes which attains unprecedented speed-up and efficiency.  The
method is based on a rectangular domain decomposition strategy.  Similar
partitioning techniques have been used by Tucker,$^{18}$ Embrechts {\it et
al},$^{22}$, Baillie and Coddington$^{21}$ and by D.  Rapaport who has
simulated extremely large systems ($640000^2$) by sequentially loading one
square subsystem at a time on an IBM RS/6000$^{23, 20}$.  In our scheme the
clusters are first labeled locally inside each processing node and then
labels are propagated across processing nodes by a relaxation process. In
section two we describe the algorithm and discuss some of its properties.
In section three we will analyze its time complexity and scaling
properties.  Numerical results are discussed in section four.  Finally,
section five contains conclusions.

\midinsert\vskip 9cm
{\rightskip=1.5pc \leftskip=1.5pc \eightrm\baselineskip=10pt\noindent
Fig. 2.\,\,\, (a) shows the initial state
and connectivity of a sub-system on one processing node. (b) shows the
local roots (in black) and the sites that point to them. (c) illustrates
the data structures setup for the relaxation cycle.}
\endinsert

\vglue 12pt
\line{\tenbf 2. Description of the Algorithm\hfil}
\vglue 5pt

Physical space is divided into rectangular cells in such way that each cell
is assigned to one processing node (see Figure 1). The algorithm labels the
clusters in two stages: first it finds all the clusters inside each
processing node using a serial local algorithm, and then it performs a global
relaxation process in which processing nodes exchange clusters labels
with nearest neighbors until a fixed point is reached. The operations of
the algorithm are shown in Figures 2 and 3. The procedure can be described
as follows:

\medskip

\vglue 0.35cm
{\tenbf Procedure Cluster-Labeling:}\quad
\medskip
\halign{\indent#\hfil&#\hfil \cr

(i) &Define connectivity for the sites: for Swendsen-Wang dynamics
connectivity\cr
    &bonds are defined with probability $p_{bond} = 1 - e^{-2\beta}$
between aligned spins.\cr

(ii) &In each processing node independently apply the serial algorithm to
find\cr
     &clusters (Procedure Local). At boundary sites the off-node bonds are
ignored.\cr
     &At the end all sites are labeled with their corresponding ``local
roots'',\cr
     & which are then globalized\footnote"$^{v}$"{\eightrm\baselineskip=10pt
i.e. made unique
over the whole system. This is done by adding an offset equal to the
processing node number times the size of the local system ($n$). For the
Swendsen-Wang  dynamics we also multiply the result by two and add a random
bit -- in this way we can ``piggy-back'' flipping information in the parity
of the cluster label.}.\cr

(iii) &Iterate relaxation cycles, exchanging local root labels with
neighboring\cr
      &processing nodes until there is no change in any node (Procedure
Relax).\cr
      & At the end all sites get their final global label from their local
roots.\cr

(iv) &Clusters are flipped with 50\% probability and measurements of
relevant\cr
     &quantities are accumulated (energy, magnetization etc.).\cr
}

\vglue 0.35cm

\midinsert\vskip 9cm
{\rightskip=1.5pc \leftskip=1.5pc \eightrm\baselineskip=10pt\noindent
Fig. 3.\,\,\, Processing nodes exchange
local root labels with nearest neighbors. Minimum labels are found and
local root labels are updated. The process is repeated until there are no
more changes in local root labels.}
\endinsert

The serial local algorithm we used has similarities to both
Hoshen-Kopelman$^{24}$ and union-find$^{10}$
algorithms\footnote"$^{vi}$"{\eightrm\baselineskip=10pt We also used some
ideas from an earlier serial code developed by R. Giles, R.  Brower and P.
Tamayo at Boston University.}. This particular serial algorithm proved
efficient for our approach but other serial algorithms can be used.  It can
be briefly described as follows:

\vglue 0.35cm
{\tenbf Procedure Local:}\quad
\medskip
\halign{\indent#\hfil&#\hfil \cr
(i) &Assign a unique label to each site (e.g. if sites are stored in a one \cr
    &dimensional array then the label is the array index of the site
itself).\cr

(ii) &For each site, follow the label paths for the site and each of its
connected\cr
     &neighbors until their roots (sites pointing to themselves) are found.
Then\cr
     &obtain the minimum label and set the site label, its root, and the roots
of\cr
     &the connected neighbors to the minimum value.\cr

(iii) &To finish the labeling process a final collapse of trees is done by a
pass\cr
      &over all sites setting each site to point to its corresponding root.\cr
}

\vglue 0.35cm

After the local labeling is completed (see Fig. 2b) a number of relaxation
cycles are executed to label the clusters globally. The relaxation
procedure consists of the following:

\vglue 0.35cm
{\tenbf Procedure Relax:}\quad
\medskip
\halign{\indent#\hfil&#\hfil \cr
(i) &Execute a preparation step to set data structures: for each node
boundary\cr
    &define a list of pointers to local roots (see Fig. 2c), one pointer per
each bond\cr
    &crossing node boundaries.\cr

(ii) &Execute a sequence of relaxation cycles until all nodes detect no
change\cr
     &in the labels (see Fig. 3): \cr

{} {} {} {} (a) &{} {} Each processing node interchanges boundary labels with
the neighboring\cr
       &{} {} node in each direction. The labels are sent in a single block of
data using\cr
       &{} {} synchronous message-passing calls.\cr

{} {} {} {} (b) &{} {} The local root labels are compared with the ones
obtained from the\cr
       &{} {} neighbors and then set to the minimum values.\cr
}
\vglue 0.35cm

We find that even for large numbers of processing nodes the execution time is
dominated by the local part. Originally we intended to implement a
multi-grid scheme for the relaxation part, as was used in ref. 25.  This
extension is straightforward and makes the algorithm more scalable but we
have found that in practice it is not necessary given the sizes of
coarse-grained MIMD machines available today. Even a large parallel
computer with 1024 processing nodes forms relatively small lattices in 2
and 3 dimensions which would require only a few multigrid levels.
Nearest-neighbor relaxation is very efficient for this problem at the scale
of the processing nodes grid.

\vglue 12pt
\line{\tenbf 3. Analysis: Complexity and Scaling\hfil}
\vglue 5pt

Considering the fact that the cluster labeling algorithm operates at the
core of equilibrium Monte Carlo simulations the analysis will focus on
average case instead of worst case performance. The probability of
obtaining a given configuration of clusters is determined by the Boltzmann
weight of that particular configuration. The probability of observing the
worst case is negligible. The execution times we will consider correspond
to averages over equilibrium configurations generated in the Monte Carlo
process. Furthermore, we will analyze the properties of the algorithm at
the critical point for the Ising model.

In this section we will make a simplified model for the time complexity of the
algorithm based on simple scaling arguments. The predictions of the model
will be compared with experimental results in the next section. Similar
analysis for the geometric parallelization method have been presented by
Burkitt and Heermann$^{19}$ and Jakobs and Gerling$^{26}$.

In the following discussion, $N = L \times L$ is the total size of the
system, $n = l \times l$ is the size for the subsystem in each processing
node, and $p$ is the total number of processing nodes. Clearly the total number
of
sites is $N = np$.  In addition, $a$ and $b$ are constants. We start by
considering the total time to perform the cluster labeling as consisting
of two contributions: a ``local time'', the time spent by the serial
algorithm inside each processing node, and a ``relax time'' which is the
time spent in the global relaxation procedure until completion.  The
scaling of the local part is basically $O(n\; log^*n)$ where $log^*n$ is a
very slow growing function\footnote"$^{vii}$"{\eightrm\baselineskip=10pt
$\log^*n$ is equal to the number of times the $\log$
function has to be applied recursively to the argument until it converges
to one.}. In practice, as it is discussed in ref. 10, the value of $log^*n$
can be safely considered a constant and the serial time is basically
$O(n)$,

$$ T_{local} \; = \; a l^2 \; = \; a n. \,\, \tag 1$$

To compute the relaxation time we need to compute two contributions:
$t_{relax}$, the time to do one relaxation cycle and $n_{relax}$, the number of
relaxation cycles needed to complete the process. $t_{relax}$ is proportional
to
the size of the boundary between processing nodes, therefore, assuming the
communication times grow linearly with message size, we have,

$$ t_{relax} \; \sim \; l \; \sim \; n^{1/2}. \,\, \tag 2$$

If we consider relaxation as a nearest-neighbor propagation of labels, then
$n_{relax}$ should be proportional to the maximum depth of the clusters
embedded in the lattice. The maximum depth of the clusters is equivalent to
the maximum shortest path joining two sites over the set of clusters:
$\lambda_{min}$; this length, also known as the chemical distance$^{27}$,
characterizes the way information is transmitted inside a cluster by
step-by-step processes such as nearest neighbor label propagation.
Typically for percolation clusters, $\lambda_{min}$ scales with cluster
linear size $r$ with a characteristic exponent $d_{min}$,

$$ \lambda_{min} \; \sim \; r^{d_{min}}, \,\, \tag 3$$

\noindent
which implies that the number of relaxation cycles to label a given cluster
will scale with $d_{min}$. Since the clusters can be as large as the entire
system $(r = L)$, then $n_{relax}$ will be proportional to $L^{d_{min}}$.
However, in our case we have relaxation only at the scale of the processing
nodes, not at the scale of single sites, and then the correct length to use
in the scaling expression for $n_{relax}$ is the renormalized length $L/l =
p^{1/2}$,

$$ n_{relax} \; \sim \; (L/l)^{d_{min}} \; \sim \; p^{d_{min}/2}. \,\, \tag
4$$

The connectivity at lower scales is ``integrated out'' by the local procedure
but we are assuming that the chemical distance of the resulting
renormalized lattice still scales with the same exponent $d_{min}$. The
value of $d_{min}$ for 2d Coniglio-Klein clusters is reported to be$^{28}$
$1.08 \pm 0.01$; consequently we expect $n_{relax}$ to grow slightly faster
than the square root of the number of
processing nodes\footnote"$^{viii}$"{\eightrm\baselineskip=10pt The use of a
multigrid or hierarchical scheme can make $n_{relax}$ scale as $\log^2 p$,
see for example refs. 25 and 29.}. Now we can express the total relaxation
time, $T_{relax}$, as

$$ T_{relax} \; = \; n_{relax}\;  t_{relax} \; = \; b p^{d_{min}/2} n^{1/2},
\,\, \tag
5$$
\noindent
where we can see that $d_{min}$ plays the role of a ``computational''
critical slowing down exponent for the algorithm. This is one of the cases
in which a physical parameter of the simulated system, such as the chemical
length exponent, determines its computational properties.

The total time for the parallel algorithm is then the sum of local plus relax
times,

$$ T_{parallel} \; = \; T_{local} + T_{relax}\; = \; a n + b p^{d_{min}/2}
n^{1/2}. \,\, \tag 6$$

If $n$ is large compared with $p$, and the communication to computation
ratio $a/b$ is small, the scaling will be dominated by the local part. This
can be seen more explicitly by calculating the speed-up function $S$, which
is the ratio between the serial and parallel times,

$$ S \; = \; { T_{serial} \over T_{parallel}}. \,\, \tag 7$$

We will use as $T_{serial}$ the time of the local algorithm running on only
one processing node.  We can express $S$ as a function of $n$ and $p$ in
the following way,

$$ S(n, p)  \; = \; { anp \over an + b p^{d_{min}/2} n^{1/2}}. \,\, \tag 8$$

As expected the speed-up improves with large $n$ and gets worse as $p$
increases. Usually, the speed-up is computed as a function of $p$ for fixed
system size $N$,

$$ S_N(p) \; = \; { aNp \over aN + b p^{(d_{min} + 1)/2}
N^{1/2}}. \,\, \tag 9$$

The corresponding efficiency $E_N(p) = S_N(p)/p$ is

$$ E_N(p) \; = \; { 1 \over 1 + {b \over a} [p^{(d_{min} + 1)} N^{-1}]^{1/2}}.
\,\, \tag 10$$

\midinsert\vskip 9cm
{\rightskip=1.5pc \leftskip=1.5pc \eightrm\baselineskip=10pt\noindent
Fig. 4.\,\,\, Local, relax and total times for
2d SW dynamics as a function of $N^{1/2}$ for fixed $p = 256$.}
\endinsert

\midinsert\vskip 9cm
{\rightskip=1.5pc \leftskip=1.5pc \eightrm\baselineskip=10pt\noindent
Fig. 5.\,\,\, Local, relax and total times for
2d SW dynamics as a function of $p$ for fixed $n$.}
\endinsert

The efficiency decreases as the number of processing nodes increases because
inter-processor communication times will eventually dominate. In practice it is
important to make $b/a$ as small as possible. As we will see in
the next section, this can be done effectively on the CM-5 where $b/a
\simeq 1$. It is interesting to notice that the efficiency is a universal
function of $p^{(d_{min}+1)} N^{-1}$. In terms of efficiency the only
important parameter in the simulation is the value of
$p^{(d_{min}+1)}N^{-1}$: the larger $N$ is in relation to $p$ the better the
algorithm will perform.

\vglue 12pt
\line{\tenbf 4. Numerical Results\hfil}
\vglue 5pt

We have implemented the algorithm using standard C language plus
message-passing calls (for inter-processor communication) using the CM-5
CMMD library\footnote"$^{viii}$"{\eightrm\baselineskip=10pt The program is
about 600 lines of code and it will be available from the authors.}.
Detailed information about the CM-5 and its network architecture can be
found in ref. 30.

\midinsert\vskip 9cm
{\rightskip=1.5pc \leftskip=1.5pc \eightrm\baselineskip=10pt\noindent
Fig. 6.\,\,\, The number of relaxation
cycles as a function of the number of processing nodes for fixed $n=128^2$. The
dot-dashed line gives the asymptotic slope predicted by $d_{min}/2$.}
\endinsert

We have performed simulations of the 2d Ising model with Swendsen-Wang
dynamics at the critical point and measured execution times for different
values of $n$ and $p$, and total system sizes $N$ from $256^2$ to
$27808^2$, using up to $256$ processing nodes on CM-5 machines without
vector units (SPARC node processors). Tables 1 and 2, and Figures 4 and 5
show times (local, relax, and total) for different system sizes and numbers
of processing nodes.  Measurements times (energy, magnetization, etc.) were
not included in the timings. As we can see in Figures 4 and 5, local times
dominate for large systems and we obtain good performance. Typical updating
times are $314$ nanosecs/site for a $64$ node CM-5 and $82$ nanosecs/site
on a $256$ node CM-5.

\midinsert\vskip 9cm
\centerline{\eightrm\baselineskip=8pt Fig. 7.\,\,\, Total time as a function of
inverse temperature $\beta$.}
\endinsert

\midinsert\vskip 9cm
\centerline{\eightrm\baselineskip=8pt Fig. 8.\,\,\, Speed-up data for up to 256
processing nodes for different system sizes.}
\endinsert

\bigskip
\bigskip
\centerline{\eightrm Table 1. Timings for 2d SW dynamics.}
$$\vbox{\eightrm\halign{\hfil#\quad\hfil &#\hfil\quad
 &\hfil#\hfil\quad
 &\hfil#\hfil\quad &#\hfil\quad &#\hfil\cr
\multispan5\hrulefill\cr
&\hfil \hfil &\hfil $p$ = 64\hfil &\hfil \hfil &\hfil \hfil\cr
\multispan5\hrulefill\cr
$N$ &\hfil local [secs] \hfil &\hfil relax [secs] \hfil &\hfil
total [secs] \hfil &\hfil nanosecs/site\hfil\cr
\multispan5\hrulefill\cr
256$^2$ &  0.019  &  0.010  &  0.029  & 442\cr
512$^2$ &  0.076  &  0.014  &  0.090  & 343\cr
1024$^2$ & 0.307  &  0.021  &  0.328  & 313\cr
2048$^2$ & 1.264  &  0.027  &  1.291  & 308\cr
4096$^2$ & 5.188  &  0.089  &  5.277  & 314\cr
8192$^2$ & 20.94  &  0.143  &  21.09  & 314\cr
\multispan5\hrulefill\cr}}$$

\smallskip
\centerline{\eightrm Table 2. Timings for 2d SW dynamics.}
$$\vbox{\eightrm\halign{\hfil#\quad\hfil &#\hfil\quad
 &\hfil#\hfil\quad
 &\hfil#\hfil\quad &#\hfil\quad &#\hfil\cr
\multispan5\hrulefill\cr
&\hfil \hfil &\hfil $p$ = 256\hfil &\hfil \hfil &\hfil \hfil\cr
\multispan5\hrulefill\cr
$N$ &\hfil local [secs] \hfil &\hfil relax [secs] \hfil &\hfil
total [secs] \hfil &\hfil nanosecs/site\hfil\cr
\multispan5\hrulefill\cr
512$^2$ &    0.020  &  0.025  &  0.045  &  172 \cr
1024$^2$ &   0.079  &  0.027  &  0.106  &  101 \cr
2048$^2$ &   0.318  &  0.035  &  0.353  &   84 \cr
4096$^2$ &   1.277  &  0.078  &  1.355  &   81 \cr
8192$^2$ &   5.232  &  0.159  &  5.391  &   80 \cr
16384$^2$ &  21.27  &  0.341  &  21.61  &   81 \cr
27808$^2$ &  62.87  &  0.566  &  63.44  &   82 \cr
\multispan5\hrulefill\cr}}$$

Simulating a $27808^2$ system requires about $15$ Mbytes of local memory on
a $256$ node CM-5. This is about 5 bytes per site (4 bytes, one integer,
for the label and one byte for spin and bond connectivity information).
Using $32$ Mbytes of memory systems as large as $40600^2$ can be simulated
on $256$ nodes.  The local procedure is not particularly amenable to
vectorization but the use of the parallelism provided by the $4$ vector
units on each CM-5 node, which support integer operations, will increase
the speed of the local procedure significantly.

\midinsert\vskip 9cm
\centerline{\eightrm\baselineskip=8pt Fig. 9.\,\,\, Efficiency data for up to
256 processing nodes for different system sizes.}
\endinsert

\midinsert\vskip 9cm
{\eightrm\baselineskip=8pt Fig. 10.\,\,\, Universal scaling of
efficiency data plotted as a function of $[p^{(d_{min}+1)}N^{-1}]^{1/2}$.
The solid line is the predicted behavior given by Equation 10 (with $a/b =
1$).}
\endinsert

 The scaling behavior of the measured times agrees well with the simple
scaling model of the previous section (Equations 1-5): local times scale
linearly with $n$, and relaxation times with $n^{1/2}$ (see Fig. 4). The
number of relaxation cycles as a function of $p$ for fixed $n$ approaches
the asymptotic slope $d_{min}/2$ as can be seen in Fig. 6. From these data
we obtain $b/a \simeq 1$.  Figure 7 shows the total time as a function of
temperature where the peak corresponds to the critical point.

The speed-up $S_N(p)$ as a function of the number of processing nodes for
different system sizes $N$ is shown in Figure 8. Notice that for large
system sizes the speed-up keeps increasing without saturation even for our
maximum $p$ equal to $256$ processing nodes. The algorithm has much better
speed-up and efficiency than previous
methods\footnote"$^i$"{\eightrm\baselineskip=10pt For example compare our
Fig. 8 with Fig. 5 of ref. 19 (first paper) or Figures 4, 5 and 6 of ref.
21.}. The $27808^2$ lattice is simulated with 92 \% efficiency. The
functional form of these curves is given by Eq. 9. The efficiency $E_N(p) =
S_N(p)/p$ as a function of $p$ is shown in Fig. 9.  Simulations of large
systems ($ > 2048^2$) attain efficiencies higher than 90\% even for up to
$256$ processing nodes.  When these points are plotted as a function of
$[p^{(d_{min}+1)}N^{-1}]^{1/2}$ (see Fig. 10) they show the scaling
described by Eq. 10: data for different system sizes collapses reasonably
well to a universal curve. In this plot, as well as in the others, there
are fluctuations in the timings that cause small deviations from the expected
behavior. This is due to cache memory effects and non-linear behavior in
the message-passing communications. We have not tried to take these effects
into account in the scaling model because they are hard to estimate and
will make the analysis unnecessarily complicated.

\vglue 12pt
\line{\tenbf 5. Conclusions\hfil}
\vglue 5pt

We have found that rectangular domain decomposition and a combination of a
local method with global relaxation produces much better scaling properties
than other methods. A careful choice of data structures, partitioning and
communication strategies is fundamental to produce practical and efficient
algorithms for Monte Carlo simulations. A simple scaling model for the
complexity of the algorithm gives reasonable predictions for the observed
times, speed-up and efficiencies.

 Our algorithm running on a 256 node CM-5 is about 79 times faster than a
Swendsen-Wang program$^{32}$ on a Cray-XMP (6.5 $\mu$secs/site), which
clearly indicates that MIMD parallel computers, and the CM-5 in particular,
can be used very effectively for this kind of computational problems.

It is interesting to notice that many of the most efficient cluster
algorithms, SIMD and MIMD, employ relaxation or relaxation plus multigrid
methods: Brower {\it et al}$^{25}$ (1.5 $\mu$sec/site, 64K CM-2), Baillie
and Coddington$^{21,31}$ (1 $\mu$sec/site, Symult 192 nodes and nCUBE-2 64
nodes), Apostolakis {\it et al}$^{29,31}$ (6.5 $\mu$secs/site, 16K CM-2)
and this work (82 nanosecs/site, CM-5 256 nodes).  Other methods are also
being improved. Recently, Apostolakis, Coddington and Marinari$^{31}$
obtained speeds of 1.36 $\mu$secs/site on a 32K CM-2 using a global
get/send method. Kertesz and Stauffer$^{20}$ attained speeds of about 0.5
$\mu$secs/site with their Intel iPSC/860 (8 nodes) implementation of the
strip geometric parallelization method.

We hope our cluster labeling method will provide a useful tool for the
study of large systems and the calculation of critical exponents,
correlation functions etc, with unprecedented accuracy.  It can be applied
to simulations of Ising and Potts models,$^{1,2}$ embedded dynamics
simulations of Landau-Ginzburg and Heisenberg models,$^2$ virtual bond
percolation dynamics$^{33}$ and the study of static and dynamic properties
of percolation clusters$^{4,6}$.

\vglue 12pt
\line{\tenbf Acknowledgments\hfil}
\vglue 5pt

We want to thank L. Tucker, C. Feynmann, D. Stauffer, G. Blelloch, A.
Greenberg, R. Giles, R.  Brower, W. Klein, N. Copty, B. Boghosian, G.
Batrouni, R. Jones, M. Drumheller, G. Drescher, R. Frye, M. Best, P.
Coddington, J. Apostolakis and E. Marinari, for interesting discussions
about cluster methods and their applications. We want to thank one
anonymous referee for all his valuable and helpful comments which motivated
us to improve our program. We also thank the members of the CM-5 CMMD team
for their technical advice and useful suggestions. We express our
appreciation to C.  Madsen, J.  Mesirov, L.  Tucker and J.  Mucci for
supporting this project.  Some of the simulations were done on CM-5
machines at National Center for Supercomputer Applications (NCSA) and the
Pittsburgh Supercomputer Center (PSC).

\vglue 12pt
\line{\tenbf References\hfil}
\vglue 5pt

\medskip
\ninerm
\baselineskip=11pt
\frenchspacing

\item{1.} R. Swendsen and J. S. Wang,  {\nineit Phys. Rev. Lett.}, {\ninebf
58,}  (1987) 86.

\item{2.} R. Brower and P. Tamayo, {\nineit Phys. Rev. Lett.}, {\ninebf
62},  (1989) 1087; R. Brower and S. Huang, {\nineit Phys. Rev. D } {\ninebf
41},  (1990) 708; U. Wolff, {\nineit Phys. Rev. Lett.} {\ninebf 60},
(1988) 1461; R.  Edwards and A. Sokal, {\nineit Phys.  Rev. D} {\ninebf 38},
 (1988) 2009; W.  Klein, T. Ray and P. Tamayo, {\nineit Phys.  Rev. Lett.},
{\ninebf 62},  (1989) 163; P.  Tamayo, R. C.  Brower and W. Klein, {\nineit J.
of
Stat. Phys.}, {\ninebf 58},  (1990) 1083;

\item{3.} P. C. Hohenberg and B. Halperin, {\nineit Rev. of Mod. Phys.},
{\ninebf 49},  (1977) 435.

\item{4.} D. Stauffer and A. Aharony, {\nineit Introduction to Percolation
Theory 2nd. ed.,} Taylor and Francis, London (1992).

\item{5.} T. H. Cormen, C. E. Leiserson and R. L. Rivest, {\nineit Introduction
to
Algorithms}, MIT Press, (1990); D. Knuth, {\nineit The Art of Computer
Programming,} vol 3, Addison-Wesley (1973).

\item{6.} A. Coniglio and W. Klein, {\nineit J. Phys. A,} {\ninebf 13},
(1980)   2775.

\item{7.} D. Stauffer, {\nineit Physica A} {\ninebf 171},  (1991) 471; D.
Stauffer and J. Kertesz,
{\nineit Physica A} {\ninebf 177}, (1991) 381; E. N. Miranda, {\nineit
Physica A} {\ninebf 175}, (1991) 235 , {\ninebf 179}, (1991) 340. E. N.
Miranda, {\nineit Physica A} {\ninebf 175}, (1991) 229.

\item{8.} A. Coniglio, {\nineit Phys. Rev. Lett.}, {\ninebf 62},  (1989) 3054.

\item{9.} J. Woo and S. Sahni, {\nineit Jour. of Supercomputing} {\ninebf
3},  (1989) 209.

\item{10.} E. Horowitz and S. Sahni, {\nineit Fundamentals of Computers
Algorithms}, Potomac, Md. Computer Science Press, 1978.

\item{11.} M. J. Quinn and N. Deo, {\nineit Computing Surveys,} Vol. 16, No 3,
September 1984.

\item{12.} Y. Shiloach and U. Vishkin, {\nineit Jour. of Algorithms}
{\ninebf 3}, (1982)  57.

\item{13.} U. Vishkin, {\nineit Discrete Applied Mathematics} {\ninebf 9},
(1984) 197.

\item{14.} P. S. Gopalakrishnan, I. V. Ramakrishnan and L. N. Kanal, 1985
{\nineit IEEE, Int. Conf. on Parallel Processing.}

\item{15.} G. H. Blelloch, {\nineit Vector Models for Data-Parallel
Computing},  MIT Press, 1990.

\item{16.} R. Cypher, J. L. C. Sanz and L. Snyder, {\nineit Journal of
Algorithms} {\ninebf 10}, (1989) 140.

\item{17.} W. Lim, A. Agrawal and L. Nekludova, {\nineit Thinking Machines
Tech. Report} NA86-2.

\item{18.} L. W. Tucker, Proc. {\nineit IEEE Conference on Computer Vision and
Pattern
Recognition,} June 1986, Miami, Florida.

\item{19.} A. N. Burkitt and D. W. Heermann, {\nineit Comp. Phys. Comm.}
{\ninebf 54}, (1989) 210; D. W. Heermann and A. N. Burkitt, {\it Parallel
Algorithms in Computational Science}, Springer Verlag, Heidelberg 1991.

\item{20.} J. Kertesz and D. Stauffer, private communication.

\item{21.}  C. F. Baillie and P. D. Coddington, {\nineit Concurrency: Practice
and
Experience} {\ninebf 3}(2), (1991)  129.

\item{22.} H. Embrechts, D. Roose and P. Wambacq, {\nineit Hypercube and
Distributed
Computers,} F. Andre and J. P. Verjus eds.  Elsevier Science Pubs. B. V.
(North-Holland) 1989.

\item{23.} D. C. Rapaport, {\nineit  J. Phys. A} {\ninebf 18}, (1985) L175.

\item{24.}  J. Hoshen and R. Kopelman, {\nineit Phys. Rev. B,} {\ninebf
14}, (1976)  3438.

\item{25.} R. C. Brower, P. Tamayo and B. York, {\nineit Jour. of Stat.
Phys.} {\ninebf 63}, (1991)  73.

\item{26.} A. Jakobs and R. Gerling, {\nineit  Physica A} {\ninebf 180},
(1992) 407.

\item{27.} H. J. Herrmann and H. E. Stanley, {\nineit J. Phys. A} {\ninebf
21}  (1988) L829.

\item{28.} E. N. Miranda, {\nineit Physica A} {\ninebf 175},  (1991) 229.

\item{29.}  J. Apostolakis, P. Coddington and E. Marinari, {\nineit
Europhys. Lett.} {\ninebf 17}(3)  (1992) 198;

\item{30.}  C. H. Leiserson {\nineit et al}, {\nineit The Network Architecture
of
the Connection Machine CM-5}, The 1992 Symposium on Parallel Algorithms
and Architectures, San Diego, CA 1992; {\nineit The Connection Machine CM-5
Technical Summary}, October 1991. Thinking Machines Corp. Cambridge, MA.

\item{31.}  P. Coddington, J. Apostolakis and E. Marinari, private
communication.

\item{32.}  U. Wolff, {\nineit Phys. Lett. B} {\ninebf 228}, (1989)  379 .

\item{33.}  R. Brower and P. Tamayo, Virtual Bond Percolation for Ising Cluster
Dynamics, {\nineit Thinking Machines Tech. Report} TMC 231 (1992); to be
published in {\nineit Physica A}.

\vfil\supereject
\bye